\documentclass[draft]{agujournal2019}
\usepackage{url}
\usepackage{soul}
\usepackage[finalnew]{trackchanges}
\usepackage{appendix}
\usepackage{changes}   

\addeditor{rev1}
\addeditor{rev2}
\addeditor{rev3}
\addeditor{languagecorrection}

\journalname{Earth and Space Science}

\begin{document}

\title{Analysis of 42 years of Cosmic Ray Measurements by the Neutron Monitor at Lomnický štít Observatory}

\authors{Imre Kisvárdai\affil{1,2,3,4}, Filip Štempel\affil{1}, Lukáš Randuška\affil{1}, Šimon Mackovjak\affil{1}, Ronald Langer\affil{1}, Igor Strh\'{a}rsk\'{y}\affil{1}, J\'{a}n Kuban\v{c}\'{a}k\affil{1}}

\affiliation{1}{Department of Space Physics, Institute of Experimental Physics, Slovak Academy of Sciences, Ko\v{s}ice, Slovakia}
\affiliation{2}{Eötvös Loránd University, Budapest, Hungary}
\affiliation{3}{Konkoly Observatory, Research Centre for Astronomy and Earth Sciences, HUN-REN, Budapest, Hungary}
\affiliation{4}{CSFK, MTA Centre of Excellence, Budapest, Hungary}

\correspondingauthor{\v{S}imon Mackovjak, Institute of Experimental Physics SAS, Watsonova 47, 04001 Ko\v{s}ice, Slovakia}{mackovjak@saske.sk}

\begin{keypoints}
\item The longest continuous time series of neutron monitor measurements is provided and analyzed
\item There are recent TGEs reported for the first time in neutron monitor measurements
\item Neutron monitor data has the capability to be used in space weather applications
\end{keypoints}

\begin{abstract}
The correlation and physical interconnection between space weather indices and cosmic ray flux has been well-established with extensive literature on the topic. Our investigation is centered on the relationships among the solar radio flux, geomagnetic field activity, and cosmic ray flux, as observed by the Neutron Monitor at the Lomnický štít Observatory in Slovakia. We processed the raw neutron monitor data, generating the first publicly accessible dataset spanning 42 years. The curated continuous data are available in .csv format in hourly resolution from December 1981 to July 2023 and in minute resolution from January 2001 to July 2023 \cite{LMKS_NM_DATA}. Validation of this processed data was accomplished by identifying distinctive events within the dataset. As part of the selection of events for case studies, we report the discovery of TGE-s visible in the data. Applying the Pearson method for statistical analysis, we quantified the linear correlation of the datasets. Additionally, a prediction power score was computed to reveal potential non-linear relationships. Our findings demonstrate a significant anti-correlation between cosmic ray and solar radio flux with a correlation coefficient of -0.74, coupled with a positive correlation concerning geomagnetic field strength. We also found that the neutron monitor measurements correlate better with a delay of 7-21 hours applied to the geomagnetic field strength data. The correlation between these datasets is further improved when inspecting periods of extreme solar events only. Lastly, the computed prediction power score of 0.22 for neutron flux in the context of geomagnetic field strength presents exciting possibilities for developing real-time geomagnetic storm prediction models based on cosmic ray measurements.
\end{abstract}

\section*{Plain Language Summary}
The study explores the connection between space weather and cosmic rays using data from the Lomnický štít Observatory in Slovakia. We processed 42 years of Neutron monitor data, discovering distinctive events, including TGEs. Applying statistical methods, we found a strong relationship: cosmic rays and solar radio flux are significantly anticorrelated (-0.74), while there's a positive correlation with geomagnetic field strength. Notably, the neutron monitor aligns better with geomagnetic data when delayed by 7-21 hours. This correlation strengthens during extreme solar events. A prediction power score of 0.22 suggests potential for real-time geomagnetic storm prediction using cosmic ray measurements.

\section{Introduction}
\label{sec:introduction}
Cosmic ray particles provide essential information about the physical state of interplanetary space, offering a unique window into the complicated mechanisms of space weather, solar activity, and the interplanetary magnetic field, that cannot be replicated in laboratories \cite{mueller2013solar}. The statistical analysis of cosmic ray measurements not only unveils the complex nature of these high-energy particles but also holds the potential to revolutionize our understanding of potentially hazardous events affecting human astronauts and space-born electronics \cite{mewaldt1996cosmic, panasyuk2001cosmic, schwadron2010galactic}.\\

The continuous 42 years time series of data from the neutron monitor on the top of Lomnický štít (LMKS) peak  (Slovakia, 49.40°N, 20.22°E, 2,634 m a.s.l.) allow us to inspect the long-term trends in the cosmic ray flux in the energies about 3.8-30 GeV \cite{butikofer2018ground}. In addition, the 1-minute resolution data since 2001 make it possible to investigate the short-term transient events (e.g. Ground Level Enhancements (GLEs) or Thunderstorm Ground Enhancements (TGEs)) that have an influence on the neutron monitor count rates. For these reasons, we comprehensively analyzed all the data gathered by the LMKS neutron monitor using multiple approaches. First, we selected and analyzed the 12 most significant events that were captured by the LMKS neutron monitor based on extreme values of neutron count and distinct patterns of important types of phenomena. Within this list, we differentiated 3 types of events: Forbush Decreases (FDs), Ground Level Enhancements (GLEs), and Thunderstorm Ground Enhancements (TGEs). 
Another applied approach is the statistical analysis and correlation testing of the Neutron Monitor (NM) dataset with other time series related to Solar activity and Space weather. The datasets used for comparison had to be continuous for an extended period of time to achieve sufficient statistical accuracy. Other authors have already correlated neutron monitor data from the Neutron Monitor Database \cite{Steigies2009} to certain Solar, geomagnetic, and space weather indexes \cite{kudela2004cosmic, mavromichalaki2005space,  kudela2005cosmic, mavromichalaki2006space, gupta2010high, chowdhury2011solar, mandrikova2021method, dorman2021space}. However, the focus of each of these studies was to correlate measurement data of certain time intervals when specific events, like GLEs or Forbush decreases, occurred. In this work, we present the first comprehensive analysis of the full 42-year-long dataset of the LMKS neutron monitor measurements.\\
The structure of the paper is as follows. In Section \ref{sec:data} we provide and describe a corrected and continuous time series of LMKS neutron monitor count rates. Section \ref{sec:methods} contains the presentation of the types of events that we studied and the in-depth explanation of the methods of our statistical analysis of the data. In Sections \ref{sec:results} and \ref{sec:discussion} we present and discuss our results from the investigation and statistical analysis of the datasets. Section \ref{sec:conclusion} provides an overview of the whole work along with our future plans for the further development of this study.

\section{Data}
\label{sec:data}
The measurements of the cosmic ray flux on Lomnický štít (LMKS) Observatory began in 1957. Hourly data were routinely archived starting on February 1, 1968. Unfortunately, the data acquired before 1981 are not available for public access \cite{kudela2009cosmic}. The increased statistical accuracy from December 1981 with the installation of the new 8-tube NM64 neutron monitor allowed the detection of even shorter variations than 1 hour. Thus, the data time resolution of the archived data was increased to 5 minutes from 1982 and to 1 minute from 2001 \cite{kudela2009cosmic}. The detailed description of the existing instrumentation is provided by \citeA{kudela2009cosmic} and \citeA{chum2020significant}. The automatic barometric correction of the measured data was implemented since 1982. Since 2001 the data have been publicly available via web interface. The past 6 hours of data with a 1-minute resolution, the past 24 hours of data with a 5-minute resolution, and the past 30 days of data with a 1-hour resolution can be found in graphical and ASCII data format on the webpage \url{http://neutronmonitor.ta3.sk/realtime.php}. Here, the data are continuously updated and are available in real-time. Along with the current measurements the archive on the same web page includes hourly data from January 1982. The international Neutron Monitor Database (NMDB, \url{https://www.nmdb.eu/data/}) also contains data from the LMKS station, from January 2001 until June 2016 in 1-hour resolution and from June 2016 until the present in 1-minute resolution. In the NMDB Event Search Tool (NEST), the measured count rates can be retrieved in counts/s or in relative units. The percentages used for relative units presented in available data and also in this paper are normalized to the 100\% level reached in September 1986 and are equal to 1,745,200 counts per hour.\\

\begin{figure}[ht]
\noindent\includegraphics[width=\textwidth]{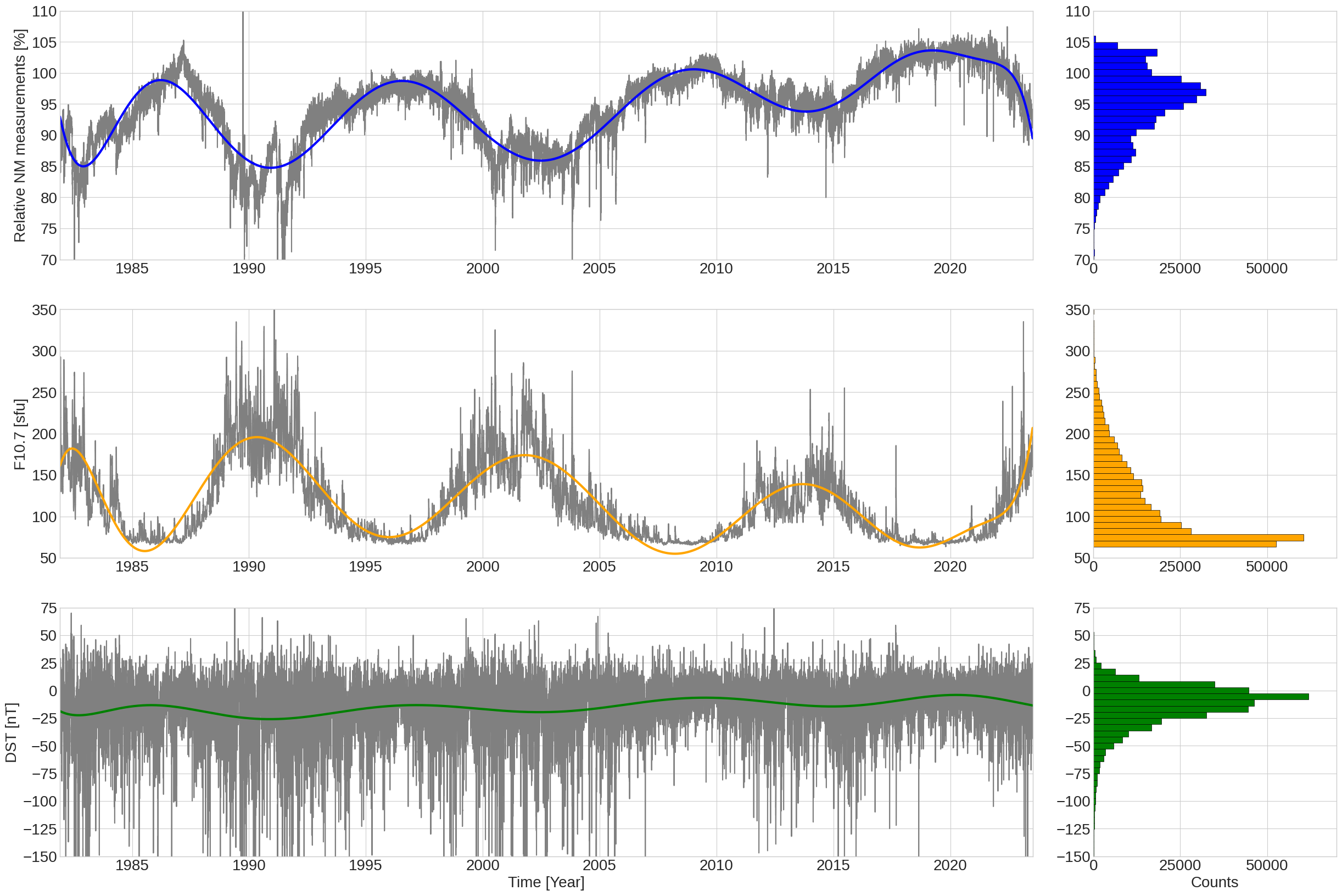}
\caption{\textit{Top panel:} The continuous time series data measured by the neutron monitor at Lomnický štít Observatory. \textit{Middle panel:} The F10.7 solar index. \textit{Bottom panel:} The Dst index. The least squares polynomial fit created using the polyfit() method from the NumPy library \cite{Harris2020} is plotted for each presented index to highlight longer periodic variability.} Histograms of all measured values can be seen on the right-hand side of the figure for each index. The purpose of color coding is only to improve visualization and distinguishing between parameters.
\label{fig:data}
\end{figure}

A continuous and curated dataset of the whole 42-year time interval, from December 1981 to July 2023,  was not available until now. (They can be downloaded now from the data repository: \url{https://zenodo.org/doi/10.5281/zenodo.10790915} \cite{LMKS_NM_DATA}). The reason was the several changes in measurement and recording techniques implemented on the measuring apparatus on LMKS. These changes and upgrades resulted in a number of different data formats throughout the years, so consolidation was required. We brought the datasets to a uniform data format and interpolated the few existing gaps in the measurement data (less than 0.016\% of the total sample intervals). It was verified that interpolation did not change data characteristics and it provides an advantage for time series usage, without gaps, in future applications. The next step was to average the neutron counts of the 4 best response characteristic tubes out of the 8 tubes that represent the LMKS neutron monitor. After we acquired the raw data of the average count rates we applied the pressure-correction function \cite{clem2000}, defined as:

\begin{equation}
\label{eq:corr}
    N_{cor} = N_{raw} e^{(\beta (p-p_m))}
\end{equation}

where $N_{cor}$ represents the pressure-corrected values of neutron counts, $N_{raw}$ is the raw data, that has not been corrected for either pressure or efficiency, $\beta$ is the barometric coefficient found by \citeA{kudela2017correlations}, $p$ is the ambient pressure value in hPa units and $p_m$ is the mean pressure at LMKS peak.\\

The visual representation of a uniform continuous time series that contains all measured data from 1981 to 2023 and that is corrected for the ambient atmospheric pressure is displayed in Figure \ref{fig:data}. The hourly data are displayed as a time series and also as a histogram to present the overall distribution. The polynomial fit of the time series is added to visualize the long-term variations. For comparison purposes, the standard space weather indexes, the F10.7 and Dst, are displayed for the same time period. The space weather indexes were downloaded from OMNIWeb \url{https://omniweb.gsfc.nasa.gov/form/dx1.html}
\cite{King2005}. The statistical analysis of these three parameters is described in Section \ref{sec:stat1} and the results are presented in Section \ref{sec:results}.

\section{Methods}
\label{sec:methods}

\subsection{Case study approach}
\label{sec:case1}

At first, we have analyzed the most significant events, by the case study approach, based on the state-of-the-art knowledge in the field. Here we provide a short overview of the literature concerning the studied categories of events presented in Section \ref{sec:results}. 

Forbush decreases (FDs) are the prevalent form of transient declines in cosmic ray (CR) flux that can be detected through terrestrial CR sensors. In these instances, the CRs are initially prevented from reaching near-Earth space by the interplanetary interferences and then it takes some time for the cosmic radiation intensity to recover. The FDs exhibit a swift reduction in intensity over a brief span of hours, followed by a gradual recovery that typically lasts several days. The amplitude of the decreases frequently amounts to a few percent or even tens of percent. These changes are recorded across multiple monitoring stations \cite{forbush1937effects, forbush1938cosmic}. For further reviews on FDs, please refer to \citeA{lockwood1971forbush, storini1990galactic, cane2000coronal, 2022SoPh..297...24L}.

Ground-level enhancements (GLEs) are unique and high-energy phenomena caused by solar energetic particles (SEPs). These events represent the uppermost energy range of SEPs, characterized by sudden and intense increases in CR intensity observed in the temporal profiles. Unlike the gradual fluctuations of the general CR background, GLEs exhibit rapid and sharp spikes of a few percent to tens of percent. These enhancements typically last for brief periods, with an initiation phase of 10 minutes or more and a slower, less intense setback phase to normal levels. What distinguishes GLEs is their occurrence at the Earth's surface level, making them a direct influence on terrestrial environments. GLE-associated SEP fluxes are often described by two distinct spectral phases: a softer phase dominated by medium-energy particles in the megaelectronvolt (MeV) range, and a harder phase where high-energy particles in the gigaelectronvolt (GeV) range dominate. On the LMKS neutron monitor data, only the harder phase of GLEs can be detected due to the cutoff rigidity of 3.85 GV \cite{belov1996proton, bieber2004latitude, kudela2009cosmic, berrilli2014relativistic, battarbee2018multi}.

Thunderstorm ground enhancements (TGEs) when detected at ground level, represent fast neutron flux increases of low intensity that extend over several minutes. TGEs are characterized by a modest escalation of neutron flux compared to the background level, typically not exceeding a few tens of percent during most TGE events \cite{chum2020significant}. However, instances of TGEs surpassing background levels by multiple factors have also been documented \cite{dwyer2014physics, chilingarian2010ground}. The initiation of TGE events is not caused by lightning, in fact, some TGEs are terminated when lightning occurs and after their termination the neutron monitor count rates return to normal levels \cite{kudela2017correlations, chum2020significant}. TGEs exhibit a correlation with elevated electric field magnitudes, suggesting their potential origin as Bremsstrahlung radiation from relativistic runaway electron avalanches (RREA), as proposed by \citeA{gurevich1992runaway}. RREA cause atmospheric cascades, that produce neutrons with near relativistic velocities. These phenomena predominantly manifest on lofty mountain summits or at lower altitudes during winter thunderstorms along the Japanese coastline \cite{chilingarian2011particle, kuroda2016observation, vslegl2022spectrometry}. For further discussion on the properties and mechanisms of TGEs, see e.g. \citeA{khaerdinov2005cosmic, mccarthy1985further, shah1985neutron, babich2014analysis, enoto2017photonuclear}.
 
\subsection{Statistical approach}
\label{sec:stat1}

Given that the analyzed dataset from the Lomnický štít neutron monitors is a 42-year-long continuous high-quality dataset, we first wanted to check the long-term linear correlations with other similar indices that are connected to space weather and solar activity. For this purpose, we chose to conduct the Pearson correlation test, as it is a widely used and precise method to measure the correlation between two independent variables. The test results in a correlation coefficient $\rho$ indicating the type and strength of the correlation present between the two variables. The value of $\rho$ ranges from -1 to +1. If $0<\rho\leq1$ the two variables have a positive linear correlation, if $\rho=0$ they are uncorrelated and for the case of $-1\leq\rho<0$ the correlation can be evaluated as a negative linear correlation \cite{cohen2009pearson}. For values $|\rho|\geq0.7$ the correlation can be considered strong \cite{schober2018correlation}.

To test whether the neutron monitor data might be suitable for creating a prediction model for the Dst index, we also conducted a predictive power score test, which is a reliable technique for discovering significant relationships between datasets due to it being data type agnostic and able to detect even non-linear relationships \cite{Wetschoreck2020}. Our reason behind searching for non-linear relationships is that the absence of a clear linear relationship is evident when plotting the Dst index values against the neutron monitor measurements. The predictive power test results in a score between 0 and 1 while 0 indicates no predictive power and 1 indicates perfect predictive power which means completely dependent datasets. Computing this score is done in multiple steps.
First, we created a regression decision tree, which requires a predictor dataset (neutron monitor data in our case) and a dataset that acts as criterion (Dst index data in our case). The decision tree algorithm (i.e. DecisionTreeRegressor() method implemented in the SciPy package \cite{Virtanen2020} and utilized according to \citeA{Wetschoreck2020}) follows a tree-like model of decisions and their possible consequences. The algorithm works by recursively splitting the data into subsets based on the most significant feature at each node of the tree. During the calculation process, we created a decision tree regressor object, to which we fitted the predictor and criterion data. We used this model to predict the criterion values again from predictor values, which returned an array of predicted criterion. In this case we were using the whole data, and not separating it into train and test datasets, because we did not want to actually predict, we just wanted to see how accurately it can describe the criterion (even if the tree has already seen all the values). For evaluation of the performance we used the normalised root mean square error, which is calculated as follows in Equation \ref{eq:nrmse}:

\begin{equation}
\label{eq:nrmse}    
nRMSE = \frac{\sqrt{\frac{1}{N} \sum_{i=1}^N (M_i - O_i)^2}}{\mbox{max}(O)-\mbox{min}(O)}
\end{equation}
where $M$ represents the decision tree model output, while $O$ corresponds to the observed values. $N$ stands for the number of samples in our data. In Equation \ref{eq:nrmse}, RMSE is normalised using the min-max method. We defined the upper limit as 0 because the perfect RMSE score is 0. The lower limit depends on the evaluation metric and dataset, which is the value that a naive predictor achieves. We can assume a prediction of the median of our data naive in all cases, therefore we can again use Eq. \ref{eq:nrmse} and replace the tree model output with the median, thus computing evaluation metric for a naive model. Finally, calculating predictive power score (PPS) is as follows in Equation \ref{eq:ppscore}:

\begin{equation}
\label{eq:ppscore}    
PPS = 1 - \frac{nRMSE_{model}}{nRMSE_{naive}}
\end{equation}
where $nRMSE_{model}$ denotes the evaluation metric of our decision tree model, while $nRMSE_{naive}$ is the evaluation metric of the naive model. Calculating this gives a value between 0 and 1, representing how well is the independent variable suitable for predicting the target variable. Results of these calculations as well as the previously mentioned correlation analysis are presented in Section \ref{sec:stat2}.

\section{Results}
\label{sec:results}


\subsection{Case study results}
\label{sec:case2}

\begin{figure}
\includegraphics[width = \textwidth]{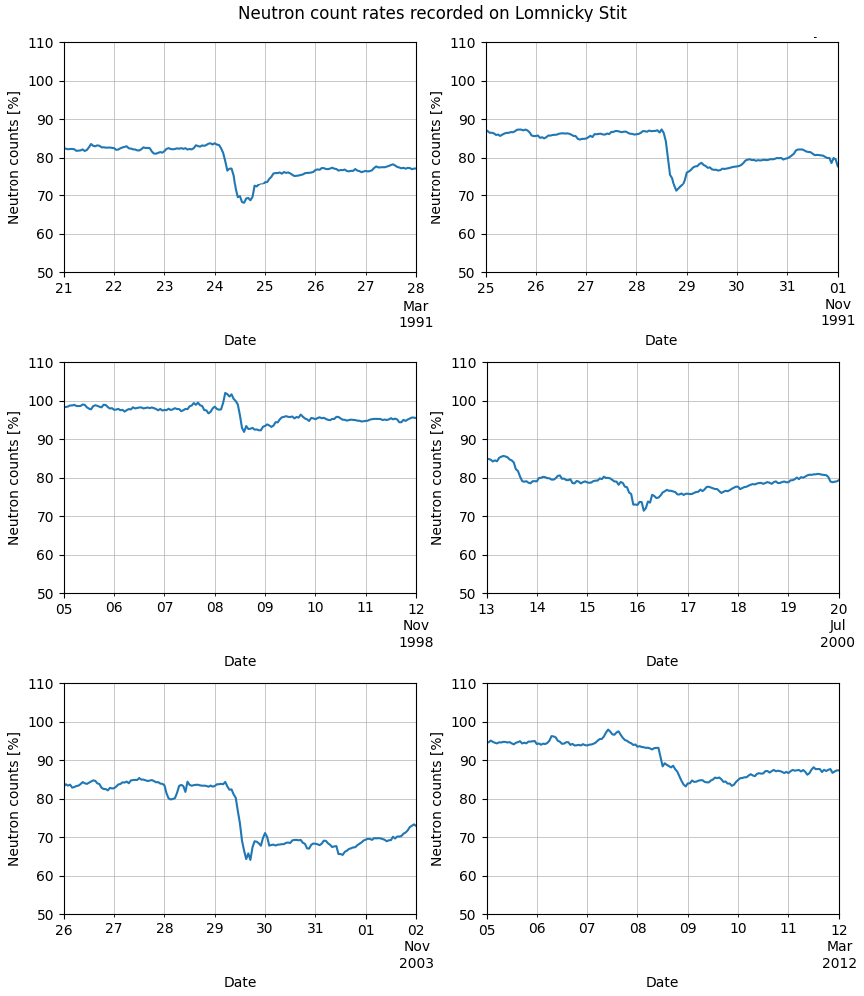}
\caption{The selected Forbush Decreases (FDs) with amplitudes ranging from 10 to 20\%. The timescale is exactly 7 days on each panel and the scale of the y-axes is also uniform. }
\label{fig:fds}
\end{figure}

\begin{figure}
\includegraphics[width = \textwidth]{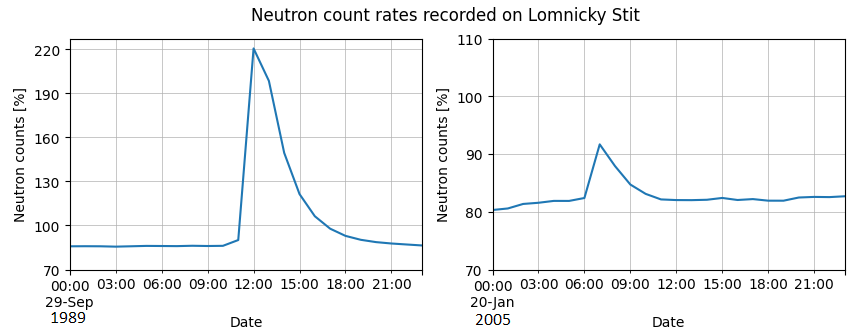}
\caption{The selected Ground Level Enhancements (GLEs). The time scale for each panel is 22 hours with a resolution of 1 hour. Note the different scaling on the y-axes.}
\label{fig:gles}
\end{figure}

\begin{figure}
\includegraphics[width = \textwidth]{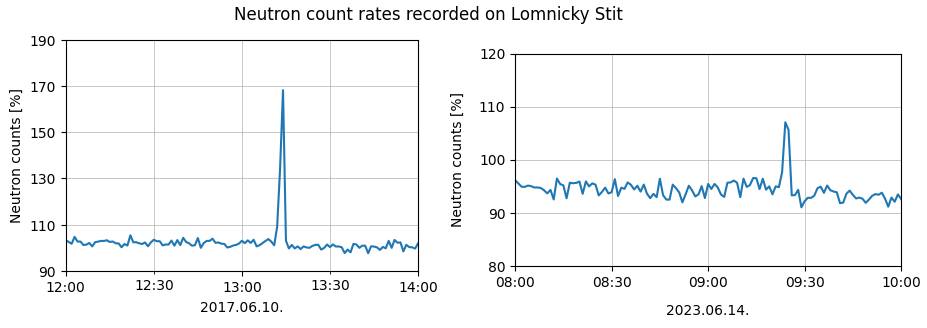}
\caption{The selected Thunderstorm Ground Enhancement (TGE) events. They are displayed on a 2-hour long time interval. The graphs are produced by minute-resolution data. The scaling of the vertical axis in the plots differs, as the first event exhibited value changes of considerably higher amplitude, than the other one.}
\label{fig:tges1}
\end{figure}

\begin{figure}
\includegraphics[width = \textwidth]{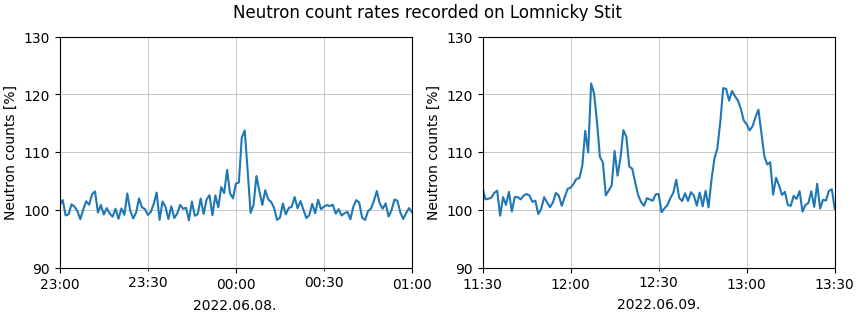}
\caption{Presumably, 3 recent TGEs reported for the first time. The data shown here is displayed on a 2-hour interval. Note the unusually small time difference between the events.}
\label{fig:tges2}
\end{figure}

In this section, we first present the selected events recorded by the neutron monitor data, that showed either characteristic features of important event types regarding their space weather related effects or reached extreme values in the neutron monitor count rates. The typical event categories are the Forbush decreases, GLEs, and TGEs described according to the existing literature in Section \ref{sec:case1}. In the following figures we showcase the most interesting events we collected from the data. It is noted that on each of Figures \ref{fig:fds}--\ref{fig:tges2} the y-axis represents the relative neutron count rates. It is just reminded that the standard for this measure is the level reached in September 1986 and represented 1,745,200 counts per hour.

In Figure \ref{fig:fds} six separate, independent events are present from a variety of time periods. Generally, the showcased events all represent typical features characterizing FD. However, each of these FDs are individually discussed in this subsection due to the varying amplitude and time span of the different phases of the events.\\

The Forbush decreases that occurred on 1991-03-24 and 1991-10-28 brought a significant change in the galactic cosmic ray flux of 15-20\% each. These are very standard cases of the phenomenon, with rapid decreases of ~10 hours and longer recovery phases lasting a few days but their amplitude is significantly higher than the average amplitude of FDs.
On 1998-11-08 the onset of the FD was preceded by a noticeable increase in the cosmic ray flux. The possible mechanisms causing this phenomenon are discussed in Section \ref{sec:discussion}. The amplitude of decrease reached 10\% and it had a more rapid recovery phase to normal levels than regular.
The panel beginning on 2000-07-13 showcases two consecutive Forbush decreases occurring with a 2.5-day delay. Both decreases had an amplitude $>5\%$ summing up to $>15\%$. This event is also further discussed in Section \ref{sec:discussion}.
The largest decrease included in this study was recorded on 2003-10-29. An intense 20\% drop in the count rates can be seen on the figure within the span of 8 hours.
The most recent event showcases an unusually extended decreasing phase of almost 1 day. The amplitude of this decrease was ~10\%.

The ground level enhancement (Figure \ref{fig:gles}) which took place on 1989-09-29 started at 10:15 and lasted more than 8 hours. The maximum value of the relative neutron counts during this period was 220\%, which is the most significant change detected compared to the base level measured before the event. The second panel of Figure \ref{fig:gles} presents a more typical GLE event. The sudden increase reached its peak value on 2005-01-20 at 07:05. Both events are confirmed and cataloged GLEs. They were observed across most neutron monitor stations associated with the Neutron Monitor Data Base (NMDB) as can be seen on the following website: \url{https://www.nmdb.eu/nest/}.

The selected Thunderstorm Ground Enhancements (TGEs) are displayed in Figure \ref{fig:tges1} and \ref{fig:tges2}. The first event strongly exhibits traits characteristic of a TGE i.e. it lasted only about 5 minutes and it was almost instantly terminated in the end.This is the largest TGE amplitude ever recorded, which exceeded the background values about 215 times in the measurements of the SEVAN (Space Environment Viewing and Analysis Network) instrument \cite{chum2020significant}. This event demonstrates the advantages of having detectors measuring the gamma-ray flux (i.e. SEVAN) and the neutron flux (i.e. Neutron monitor) simultaneously on Lomnick\'{y} \v{s}t\'{i}t. Out of all the high-altitude stations worldwide, only Lomnick\'{y} \v{s}t\'{i}t and Aragats (Armenia) host both these detectors, making them especially important places for gathering measurement data of TGE events and their interaction with atmospheric nuclei \cite{chilingarian2018sevan}. The geographic environment surrounding Lomnick\'{y} \v{s}t\'{i}t puts it in a truly unique position, where it regularly gets engulfed in thunderclouds, thus the TGEs here cause extremely high count rates for the particle detectors. Evidence of the presence of thunderstorms in the area surrounding Lomnick\'{y} \v{s}t\'{i}t can be found in the following database: \url{https://www.blitzortung.org}. An another intensive TGE event was detected by neutron monitor on 2023-06-14 at 09:25 (Figure \ref{fig:tges1}, right). Its occurrence was verified by the agreement of each neutron monitor counter tubes, the SEVAN instrument, the ambient electric field measurement system, and by the thunderstorm logs from the above-mentioned database. Therefore, we are confident in its classification as a TGE event. The classification of the June 2022 events as TGEs (Figure \ref{fig:tges2}) is not so straightforward. Enhancements of the neutron flux amplitude are not supported by the expected outstanding increases in measurements recorded by SEVAN. However, there were 20-30\% surges in the recorded flux in each three of the SEVAN channels exactly at the same time as detected by the neutron monitor. TGE events probably occurred but the possible discrepancy could be explained by the fact that during thunderstorms, the increased electric field might affect the analog part of the neutron monitor measuring system. This hypothesis is not yet verified and further investigation is necessary in future works. This case emphasizes the importance of the combination of the neutron monitor and SEVAN instruments in unveiling the source of sudden variations \cite{karapetyan2024forbush}.

\subsection{Statistical analysis results}
\label{sec:stat2}

We correlated the datasets presented in Section \ref{sec:data} with each other by leveraging the Pearson method described in Section \ref{sec:stat1} and implemented in the Pandas library \cite{Pandas2020}. The basic case of the tests we conducted was the correlation of the entirety of the datasets with no time delay between them. The Pearson coefficients resulting from this method are -0.743 and 0.306 for the F10.7 and Dst indices respectively, as shown in Table \ref{tab:pearson}. We did not conclude these statistical tests solely on the whole, continuous 42-year-long datasets, but also on time intervals capturing Forbush decreases and GLEs. A selection of Forbush decreases and GLEs were chosen from their respective databases (\url{http://spaceweather.izmiran.ru/eng/dbs.html}, \url{https://gle.oulu.fi}). We selected manually some TGE events' intervals, too. We proceeded with the correlation testing of the datasets only in the time intervals corresponding to these selections of events. Results for these tests can be seen in Table \ref{tab:pearson}. As the final step, we added the possibility of time delays between the sets as it is discussed below. It is noted, that the p-values for every coefficient in Table \ref{tab:pearson} are below 0.01, indicating that our results are significant on a 99\% confidence level.

\begin{table}
    \centering
    \begin{tabular}{lcccc}
        \hline
        & F10.7 & Dst & Delayed F10.7 & Delayed Dst \\
        \hline
        Overall neutron flux & -0.743  &  0.306 & -0.742 & 0.314 \\
        Neutron flux during FDs & -0.744 & 0.555 & -0.744 & 0.589 \\
        Neutron flux during GLEs & -0.254 & 0.340 & - & - \\
        Neutron flux during TGEs & -0.533 & -0.124 & - & - \\
        \hline
    \end{tabular}
    \caption{Correlation coefficients resulting from correlating the given datasets using the Pearson method described in Section \ref{sec:stat1}. The overall neutron flux means that we correlated the whole 42 years of measurement data with the F10.7 and Dst indexes. We also concluded the correlation test for time periods of Forbush decreases, GLEs, and TGE events, these are indicated by the FD, GLE, and TGE notations respectively. For the last two columns, the exact time delays are given and explained in the main text below.}
    \label{tab:pearson}
\end{table}

\begin{figure}
\includegraphics[width = 0.8\textwidth]{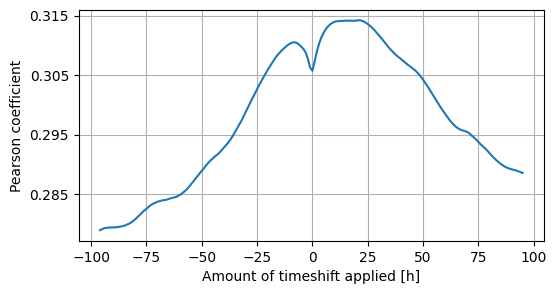}
\caption{The graph demonstrates the change of the Pearson coefficient resulting from the correlation testing of the neutron monitor and the Dst datasets by the amount of time-shift applied to the Dst data. The application of a timeshift of positive value will result in the Dst index being "brought forward in time", therefore we correlate the neutron monitor data with Dst data that were measured later.}
\label{fig:shift}
\end{figure}

Within this analysis, the F10.7 and Dst indices were shifted in one-hour increments across a range from -100 to +100 hours, to examine how potential time delays between the datasets, likely due to physical mechanisms in near-Earth space, might affect their correlation. This resulted in the maximum value of the Pearson coefficients for the neutron flux - Dst correlations of 0.314. These values are present at the interval of 7 to 21 hours of delay in the Dst data, meaning the Dst is correlated with earlier neutron monitor measurements, as presented in Figure \ref{fig:shift}. This means that the neutron monitor data is best correlated with Dst values of 7 to 21 hours in the future at times of Forbush decreases. We applied the same 7-hour time shift to the F10.7 data also, for which the results are listed in the 'Delayed F10.7' column in Table \ref{tab:pearson}.

The results of predictive power tests are presented in Table \ref{tab:pps}. These results represent the evaluation metric of our decision tree model. In the rows, the predicted indices are named and the columns contain the base indices for the prediction. The methods of these calculations are discussed in detail in Section \ref{sec:stat1}. It is noted, that we did not want to lose any valuable data from the neutron monitor measurements for the statistical analyses, therefore we interpolated the F10.7 index values as they have slightly different sampling rates.

\begin{table}
    \centering
    \begin{tabular}{cccc}
    \hline
                    & Dst & F10.7 & Neutron flux\\
    \hline
        Dst          & 1     & 0.13 & 0.22\\
        F10.7        & 0.086 & 1    & 0.49\\
        Neutron flux & 0.077 & 0.41 & 1   \\
    \hline
    \end{tabular}
    \caption{Results of predictive power test. We have conducted the predictive power test on all combinations of our data to observe, which relationships are most suitable for training a predictive model. }
    \label{tab:pps}
\end{table}

\section{Discussion}
\label{sec:discussion}
The events presented in Section \ref{sec:case2} were selected manually and verified with the method described by \citeA{OKIKE2020106463} and with the help of the world-wide neutron monitor network demonstrates that the Lomnický štít (LMKS) neutron monitor (NM) is a useful tool for measuring and analysing the common types of changes in the cosmic ray flux like Forbush decreases and GLEs. Being in the medium cutoff rigidity range (3.8 GV) the station is a valuable part of the global network of neutron monitors, since the data acquired by it helps determine the exact energy level of the particles generated during space weather events, such as Forbush decreases or GLEs. The rigidity of the particles can be further constrained by inspecting the cosmic ray flux changes recorded on NMs with higher cutoff rigidity like ROME (6.3 GV) or ATHN (8.5 GV)\cite{storini2005ground, mavromichalaki2015recent}.

The collected Forbush decreases are all confirmed cases, verified using data from multiple neutron monitor stations incorporated into the Neutron Monitor Database. Statistical correlation tests (refer to Section \ref{sec:stat2}) conducted on the cosmic ray, Dst, and F10.7 indices illustrate that during periods of Forbush decreases, the correlation between the cosmic ray measurement data and the Dst index significantly increases compared to normal periods within the datasets. The correlation with the F10.7 index shows no significant change. Additionally, we found that a 7-hour time delay between the Dst and the cosmic ray flux datasets causes another improvement in the correlation (from 0.555 to 0.589), as presented in Table \ref{tab:pearson}. This time-shift also positively impacts the general correlation between the datasets when inspected in their entirety, which can be explained by the intricate mechanisms driving the changes in the physical state and composition of interplanetary space and the magnetosphere.

The solar plasma originating from the Sun and causing the magnetic field disturbances in interplanetary space travel at velocities of 500-2000 $km/s$, while galactic cosmic ray particles typically have relativistic velocities \cite{weygand2011correlation, parker1958origin, mertsch2020test}. A consequence of this velocity difference is that the cosmic ray particles passing through the disturbed space regions reach us faster than the actual solar plasma causing the magnetic disturbance itself. Therefore, the neutron monitor measurements, sensitive to galactic cosmic rays, provide information about an onsetting magnetic storm well before it could reach Earth and affect the geomagnetic field and the Dst index. The 7-hour delay between the indices seems plausible in the context of the discussed mechanisms.\\

Prediction of geomagnetic storms is especially important for protecting the electrical components of spacecraft and the crew members of space missions, as they might not be entirely protected against incoming radiation by the geomagnetic field \cite{mavromichalaki2011applications, storini2005ground, chakraborty2020probabilistic}. Our statistical analysis of the datasets and the possible time delay between them imply that the cosmic ray flux measurements can be used to build forecasting models (utilizing e.g. machine learning techniques) to predict potentially dangerous events a few hours in advance with relatively high accuracy. Further analysis of the prediction power of the datasets and the development of actual solutions for these models are in preparation and will be presented in the follow-up publication. It is noted that the advantage of locations of neutron monitor stations in various geomagnetic latitudes (mainly at high latitudes) might leverage the prediction capabilities that were introduced here just for one station.

Looking at the results presented in Table \ref{tab:pearson}, it is no surprise that the correlation of the cosmic ray flux with the F10.7 and Dst indices did not improve when inspecting only the time periods when GLE events occurred, compared to the complete datasets. Considering the time-span of the events, we expect the increases in the values of the F10.7 index to be displaced in time by at least a few hours with respect to the increases in the neutron monitor data. This time difference causes the higher values to correspond to normal values in the other dataset, making the correlation worse. The same situation applies to the correlations during TGE events, but the reason behind the poor correlations is different. Here, we do not expect a correlation between the datasets at all since the driving mechanisms of TGEs do not influence the F10.7 and Dst indices, only the neutron monitor data, if the event is energetic enough \cite{chilingarian2020structure, kolmavsova2022continental, chilingarian2023genesis}. Further description of the background of this reasoning can be found in Section \ref{sec:case1}. Despite the mentioned lack of physical connection, the value of the correlation coefficient indicates a surprisingly strong anticorrelation at -0.533. We argue that this result is probably random, as during the correlated periods (each lasting 2 hours), the F10.7 index remains constant, since it has a resolution of 24 hours and the Pearson method can not give accurate results when we compare the varying values of the neutron monitor data with a set of constant values.

The determination of the type of the phenomena seen in Figure \ref{fig:tges2} is not a trivial problem. First, their pattern can be considered irregular in the context of TGE events, lasting longer than most recorded cases. Their proximity in time is not particularly unusual, as it is highly probable to
have several TGEs within the same storm, since the strong electric field can reside in
thunderclouds for several hours and may initiate many TGEs. The measurements of the individual tubes do not coincide very well, as only 4 out of the 8 tubes recorded significant increases in the neutron flux during the time of these events. The reason behind this might be that during thunderstorms, the increased electric field might affect analogue part of the NM's measuring system. The adjacent electric field measuring systems also displayed highly irregular patterns in their corresponding measurements. These circumstances - together with the present thunderstorm in the area at that time - let us conclude that the patterns occurring on the neutron monitors on 2022-06-08 and 2022-06-09 are probably a result of the corresponding TGE events recorded by SEVAN, despite lasting significantly longer than most documented cases. Naturally, this can easily be false the result of further investigation into these events may prove this hypothesis wrong, but if the reality aligns with our proposal, these newly discovered events increase the number of reported TGEs from Lomnick\'{y} \v{s}t\'{i}t and will add valuable samples for future analysis of TGE events. Indeed, the detailed exploration of TGEs detected at LMKS and other observatories is the content of our follow-up studies.

\section{Conclusion}
\label{sec:conclusion}

In this work, we introduced a revised dataset comprising 42 years of cosmic ray measurements conducted by the 8NM64 neutron monitor at the Lomnický štít Observatory. This dataset was constructed from raw measurement data and is now presented in a continuous and consistent format. In Section \ref{sec:results}, we explored the data to identify interesting individual events, providing a comprehensive overview of space weather phenomena captured by the neutron monitor in relation to the current understandings of the phenomena. During this phase of the study, we identified new distinct presumed thunderstorm ground enhancements (TGE) events in the data, signifying a noteworthy increase in confirmed cases detected by the Lomnický štít neutron monitor.

Additionally, we presented the results of our statistical analysis. The outcomes of these tests reveal a robust anti-correlation between the cosmic ray flux and the F10.7 data, indicated by a Pearson coefficient of -0.743. Notably, a significant positive correlation was observed for the Dst-cosmic ray flux pairing, consistent with expectations and confirmed by studies using other neutron monitor stations. We demonstrated that the correlation with the Dst index is stronger during periods of Forbush decreases, further enhanced with the application of a 7-hour time delay to the Dst index, resulting in a Pearson coefficient of 0.589 between the Dst index and the neutron monitor data during Forbush decreases.

Our calculations are consistent with current prevailing hypotheses regarding the intricate mechanisms driving space weather events and influencing the state of the interplanetary magnetic field, as discussed in Section \ref{sec:discussion}. Space weather research contributes to the understanding and resolution of various scientific and technological challenges. The complex system of phenomena within heliospheric physics remains not entirely understood; therefore, continued exploration of these topics is crucial for advancing our knowledge in this field.
The presented results carry significant scientific importance, and our work may encourage the utilization of the consistent data collected by the Lomnický štít neutron monitor, which is an important part and a valuable addition to the global neutron monitor network. Indeed, we are working on follow-up research that uses presented data as one of the input for several machine learning techniques, which performance is explored. The outputs are multi hours predictions of Dst index values. The details of this research will be provided in near future.

\section{Open Research}

The presented continuous datasets of Neutron Monitor measurements at Lomnický štít (Slovakia) are publicly available in .csv format in hourly resolution from December 1981 to July 2023 and in minute resolution from January 2001 to July 2023 \cite{LMKS_NM_DATA}. The measurements have been operated by the Institute of Experimental Physics, Slovak Academy of Sciences. We acknowledge use of NASA/GSFC's Space Physics Data Facility's OMNIWeb service and OMNI data \cite{Papitashvili_King_2020}. 

\acknowledgments
This work has been supported by Slovak VEGA research grant No. 2/0029/22. We are thankful to the experts and technicians who assured uninterrupted measurements of the Neutron Monitor instrument at LMKS over decades, namely Karel Kudela, Tibor Ďuriš, Vladimír Kollár, Slavomír Podhradský, Jozef Rojko, Jaroslav Sláma, Samuel Štefánik.
Furthermore, Imre Kisvárdai thanks the financial support provided by the undergraduate research assistant program of Konkoly Observatory and the IAU–International Visegrad Fund Mobility Awards.

\bibliography{bibliography.bib}

\end{document}